\documentclass{aa}  
\usepackage{graphicx}
\usepackage[varg]{txfonts}
\usepackage{natbib}
\bibpunct{(}{)}{;}{a}{}{,} 

\newcommand{\cobold}{{\sf CO$^5$BOLD}}
\newcommand{\LHD}{{\sf LHD}}
\newcommand{\linfor}{{\sf Linfor3D}}
\newcommand{\xx}{\ensuremath{\mathrm{1D}_{\mathrm{LHD}}}}
\newcommand{\mD}{\ensuremath{\left\langle\mathrm{3D}\right\rangle}}
\newcommand{\mlp}{\ensuremath{\alpha_{\mathrm{MLT}}}}
\newcommand{\ha}{H\ensuremath{_\alpha}}
\newcommand{\hb}{H\ensuremath{_\beta}}
\newcommand{\hg}{H\ensuremath{_\gamma}}

\newcommand{\teff}{\ensuremath{T_\mathrm{eff}}}
\newcommand{\logg}{\ensuremath{\log g}}

\newcommand{\eref}[1]{\mbox{(\ref{#1})}}
\newcommand{\ppdf}[2]{\ensuremath{\frac{\partial^2#1}{\partial #2^2}}}
\newcommand{\beq}{\begin{equation}}
\newcommand{\eeq}{\end{equation}}
\newcommand{\DIrms}{\ensuremath{\Delta_\mathrm{rms}}}
\newcommand{\sigF}{\ensuremath{\sigma_\mathrm{F}}}
\newcommand{\sigT}{\ensuremath{\sigma_\mathrm{Teff}}}
\newcommand{\chisq}{\ensuremath{\chi^2}}

\begin{document}
%

   \title{Impact of granulation effects on the use of Balmer lines as temperature indicators}


   \author{
H.-G. Ludwig \inst{1,2}
\and
N.T. Behara\inst{1,2}
\and 
M. Steffen\inst{3}
\and
P. Bonifacio\inst{1,2,4}
}

   \offprints{H.-G. Ludwig}

   \institute{
CIFIST Marie Curie Excellence Team
         \and
GEPI, Observatoire de Paris, CNRS, Universit\'e Paris Diderot; 92195
Meudon Cedex, France
\and
Astrophysikalisches Institut Potsdam, An der Sternwarte 16, D-14482 
Potsdam, Germany
\and
Istituto Nazionale di Astrofisica,
Osservatorio Astronomico di Trieste,  Via Tiepolo 11,
I-34143 Trieste, Italy
    }
\authorrunning{Ludwig et al.}
\titlerunning{3D Balmer lines}

   \date{Received ..; accepted ...}

  \abstract
  {Balmer lines serve as important indicators of stellar effective temperatures
  in late-type stellar spectra. One of their modelling uncertainties is the influence of
  convective flows on their shape.}
  {We aim to characterize the influence of convection on the wings of
Balmer lines.}
  {We perform a differential comparison of synthetic Balmer line profiles
     obtained from 3D hydrodynamical model
     atmospheres and 1D hydrostatic standard ones.
The model parameters are appropriate for F,G,K dwarf and
subgiant stars of metallicity ranging from solar to $10^{-3}$ solar.}
  {The shape of the Balmer lines predicted by 3D models can never
be exactly reproduced by a 1D model, irrespective of its effective temperature.
We introduce the concept of a 3D temperature correction, as the effective 
temperature difference
between a 3D model and a 1D model which provides the closest
match to the 3D profile. The temperature correction is different for the
different members of the Balmer series and depends on the adopted
mixing-length parameter \mlp\ in the 1D model.  Among the investigated models, the 3D correction
ranges from $-300$\, K to $+300$\, K.
Horizontal temperature fluctuations tend to reduce the 3D correction.}
  {Accurate effective temperatures cannot be derived
from the wings of Balmer lines, unless the effects of convection
are properly accounted for. The 3D models offer a physically well justified 
way of doing so. The use of 1D models treating convection
with the mixing-length theory do not appear to be suitable
for this purpose. In particular, there are indications
that it is not possible to determine a single value of \mlp\ which
will optimally reproduce the Balmer lines for any choice of
atmospheric parameters. The investigation of a more extended grid
and direct comparison with observed Balmer profiles will be 
carried out in the near future.}


   \maketitle
%

\section{Introduction}

Balmer lines 
are prominent features in stellar spectra.
There has been 
a long tradition of using the Balmer line wings
for the determination of effective temperatures
of F-K stars 
\citep[a non-exhaustive list
includes:][]{giusa1960,searle62,stro70,gehren81,cayrel1985,soderblom86}.
However, the use of Balmer lines as temperature 
indicators requires fairly high quality spectra and a sophisticated
theoretical framework, both from the point of view of micro-physics
and of the model atmospheres employed. It was not until 
the comprehensive work of \citet{F93} that the wings of
Balmer lines became the temperature indicator of choice
in many investigations. One of
the advantages of Balmer line temperatures is that, unlike those based on colours or on the 
infrared flux method, they
are reddening independent and may, in principle, provide 
an accuracy of the order of 50\,K.

In the framework of 1D homogeneous model atmospheres,
in which convective energy transport is described
by the mixing-length approximation \citep{bohm-vitense},
\citet{F93} studied in detail the effects of convection
on the wings of the Balmer lines. They
convincingly demonstrated that, as a consequence of their different
depth of formation, 
the response to a change in the adopted mixing-length 
parameter \mlp\
of the various members of the Balmer 
sequence is different. 
The first line of the series, \ha, is fairly
insensitive to the choice of \mlp, at least
for solar metallicity models, but the higher
members are fairly sensitive.
This provides, in principle, a way to select
the most appropriate value of \mlp.
For a star for which the effective temperature 
is known, like the Sun, one may select the \mlp\
which best reproduces all members of the Balmer series.

At the time \citet{F93} performed their investigation,
hydrodynamical simulations in which convection is treated
in a physical and non-parameterized way had just become
available \citep{NS91,steffen91}. However, there were not enough
simulations to cover satisfactorily the range of \teff\ and
\logg\ which was pertinent to their investigation; moreover
those simulations were not very sophisticated in 
the treatment of line opacity.
After 15 years the situation has greatly improved.
Full three dimensional hydrodynamical simulations
(hereafter 3D models, for short) and the associated line-formation
codes have reached a level of sophistication in radiative
transfer comparable to that of state-of-the art 1D models
and line-formation codes. More importantly, for the first
time a fairly large grid of 3D models is available to allow
a systematic investigation of convection effects.

To clarify how the results of \citet{F93} may be affected by inhomogeneities,
we investigate Balmer line formation in hydrodynamical model atmospheres.
Broadening due to convective velocities is only important in the inner line
cores of Balmer lines; in this sense, the outer parts of the Balmer lines are
particularly well suited sensors of the thermal structure of an atmosphere. We
use a sample of 3D model atmospheres covering the range [M/H] = --3.0 to 0.0
in metallicity, 5500-6550 K in \teff\ and 3.5-4.5 in \logg. Our investigation
is purely theoretical, and we express our results in terms of effective
temperature differences between 3D and 1D models.

\section{Model atmospheres \& line formation}

The 3D model atmospheres we use have been
computed with the \cobold\ code \citep{Freytag2002AN....323..213F,Freytag2003CO5BOLD-Manual,Wedemeyer2004A&A...414.1121W}.
The code solves the time-dependent equations
of compressible hydrodynamics coupled to radiative transfer in a constant
gravity field in a Cartesian computational domain which 
is representative of a volume located at the stellar surface.
Further details on computational methods and validity
tests can be found in  \citet{solarmodels}.
In Table \ref{tabmod} we provide some of the basic properties
of the 3D models employed in this paper.

\begin{table}
\caption{Parameters of the 3D models. The grid resolution is set to $140 \times 140
  \times 150$ ($N_\mathrm{x}\times N_\mathrm{y} \times N_\mathrm{z}$), {\em
   time} indicates the total stellar time, and {\em snaps} the number of
  snapshots per model.}             
\label{tabmod}
\centering                          
\begin{tabular}{cccccc}        
\hline
 \teff & \logg & [M/H] & box size    & time & snaps \\
   K   &  c.g.s & dex  & $x  \times y  \times z$ [Mm] &  hrs &    \\
\hline\hline
5500 & 3.5 & --2.0 & $49.0 \times 49.0 \times 35.3$ & 13.0 & 20 \\ 
5780 & 4.4 & \, 0.0 & $5.60 \times 5.60 \times 2.25$ & 1.7 & 25 \\ 
5920 & 4.5 & --3.0 & $6.02 \times 6.02 \times 3.78$ & 2.6 & 19 \\ 
6280 & 4.0 & --2.0 & $21.6 \times 21.6 \times 12.7$ & 2.7 & 16 \\ 
6320 & 4.5 & --2.0 & $7.00 \times 7.00 \times 3.95$ & 2.4 & 19 \\ 
6550 & 4.5 & --3.0 & $8.40 \times 8.40 \times 3.96$ & 0.7 & 12 \\ 
\hline
\end{tabular}
\end{table}

In order to make our comparison strictly differential
we employed as a reference 1D models computed with
the \LHD\ code, which employs the same microphysics and numerical
scheme for radiative transfer as \cobold. Further details
on these models can be found in \citet{zolfito}.
These models are classical hydrostatic 1D models; no velocity
fields are considered,
convection is treated using the \cite{Mihalas}
formulation of the mixing-length theory, 
and one has to assume a microturbulence parameter in associated spectral
synthesis calculations.
We refer to these models as \xx.

In addition we employed the 1D structure which is obtained by a temporal
and spatial average of the 3D models over surfaces of constant Rosseland
optical depth.  We refer to these models as \mD. The \mD\ model has, by
construction, the mean temperature profile of the 3D model, but no horizontal
temperature fluctuations are present. Therefore, a comparison ${\rm 3D}-\mD\ $
highlights the effects of such fluctuations, a comparison $\mD\ - \xx\ $
highlights the effect of differences in the temperature profiles, and the
3D\,--\,\xx\ comparison provides insight into the combined effects of different
temperature profiles and horizontal temperature fluctuations.


The Balmer line profiles have been computed with the \linfor\
code\footnote{http://www.aip.de/$\sim$mst/Linfor3D/linfor\_3D\_manual.pdf}.  In
the version of the code used here, the Balmer line profiles are computed with
the theory of \citet{CT60}.  In the future, more up-dated theories will be
implemented in \linfor. However, for the purpose of our strictly differential
analysis, the theory employed should not be relevant.  \linfor\ is capable of
computing line profiles both for 3D and 1D models.  The 3D synthesis is very
CPU time demanding. For this reason, we computed for each line only a range of 5.5
nm around the line center at a resolution of $1.5 \times 10^{-2}$ nm.
Furthermore, we did not include any blending lines in the wings of the Balmer
profiles. While this clearly limits the accuracy with which the higher members
of the series can reproduce the observations of more metal-rich stars, it is
not relevant for the present differential theoretical comparison.

\begin{figure}
\begin{center}
\includegraphics[width=0.45\textwidth]{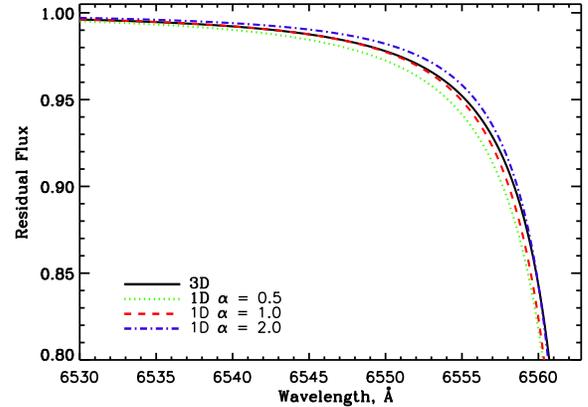}
\end{center}

\vspace{-1.5\baselineskip}

\caption[]{ \ha\ Balmer line profiles computed using a 3D model and 1D models with varying mixing-length parameter~$\alpha$, for T$_{\rm
    eff}\,=\,5920\,{\rm K}$, log g = 4.50 and [Fe/H] = --3.00. }
\label{fig1}
\end{figure}

\section{3D temperature correction}

To quantify the granulation effects, we use the concept of a 3D temperature
correction. This is defined as the difference between the effective
temperature of a 3D model and the temperature derived by fitting Balmer line
profiles computed from a \teff\ sequence of \xx\ models with the same surface
gravity and metallicity to the corresponding 3D profile.  The 3D temperature
correction is in general different for the different members of the Balmer
series (here we
investigate only the first three members) and depends on the
\mlp\ adopted for the \xx\ models.  Although we believe that the 3D temperature
correction is a useful description, one should bear in mind that it is a
simplification. In general, the whole profile computed from a 3D model has a
different shape from that of a 1D model; an example is given in
Fig.\ref{fig1}. No profile computed from a 1D model, whatever the temperature,
can {\em exactly} reproduce the profile computed from the 3D model. The
temperature correction singles out the 1D model that provides the profile {\em
  nearest} to the profile computed from the 3D model. Clearly, the concept of
distance (near or far) has to be defined by a suitable metric.

\begin{table*}
\centering
\caption{3D corrections, $\Delta$T, the quality of fit, QF, and the
  sensitivity of the fit, $\sigma_{\rm Teff}$ for the different models in
  units of K. }
\label{tbias}
\begin{tabular}{lc|clclcl|clclcl}
\hline
             &  & \multicolumn{6}{c} {3D -- 1D$_{\rm LHD}$}  &
      \multicolumn{6}{c} { $\left\langle\mathrm{3D}\right\rangle$ -- 1D$_{\rm LHD}$} \\
\hline
             &  & \multicolumn{2}{c} {\mlp\,=\,0.5} & \multicolumn{2}{c} {\mlp\,=\,1.0} & \multicolumn{2}{c} {\mlp\,=\,2.0} & 
\multicolumn{2}{c} {\mlp\,=\,0.5} & \multicolumn{2}{c} {\mlp\,=\,1.0} & \multicolumn{2}{c} {\mlp\,=\,2.0} \\
 \hline
\teff / \logg \,/ [M/H] & Line &  $\Delta$T &  QF / $\sigma$ &
$\Delta$T &  QF / $\sigma$ &
$\Delta$T &  QF / $\sigma$ & 
$\Delta$T &  QF / $\sigma$ & 
$\Delta$T &  QF / $\sigma$ & 
$\Delta$T &  QF / $\sigma$ \\
  (K / c.g.s / dex)   &      &               &           &    
  &    &             &        &      &    &   & 
  &    &  \\
 \hline \hline
5500 /3.50/--2.00  & \ha  & 235  & 89 / 26 &119   & 99 / 19 &--76   & 80 / 16 &500 & 107 / 42 &300  & 61 / 19 &85    & 60 /20 \\
                   & \hb  & 20   & 35 / 11 &--94  & 48 / 10 &--251  & 54 / 10 &256 & 22  / 14 &94   & 28 / 13 &--73  & 30 /14 \\
                   & \hg  & --10 & 31 / 10 &--144 & 44 / 8  &--309  & 51 / 9  &220 & 17  / 11 &49   & 23 / 11 &--117 & 28 /12 \\
\hline
5780/4.40/\, 0.00  & \ha  & 34   & 21 / 9  &24    & 16 / 8  &--21   & 44 / 10 &103 & 24  / 9  &94   & 14 / 9  &49    & 17 /11 \\
                   & \hb  & 39   & 15 / 9  &--23  & 52 / 9  &--164  & 72 / 9  &228 & 27  / 10 &126  & 35 / 9  &--40  & 32 /11 \\
                   & \hg  & 36   & 19 / 9  &--64  & 41 / 7  &--263  & 76 / 9  &232 & 30  / 10 &125  & 35 / 9  &--55  & 32 /10 \\
\hline
5920/4.50/--3.00   & \ha  & 285  & 65 / 16 &94    & 57 / 14 &--89   & 51 / 13 &305 & 60  / 9  &218  & 23 / 15 &21    & 35 /15 \\
                   & \hb  & 126  & 15 / 11 &--23  & 38 / 10 &--153  & 37 / 8  &235 & 14  / 11 &109  & 20 / 10 &--35  & 22 /10 \\
                   & \hg  & 105  & 18 / 9  &--57  & 26 / 8  &--179  & 40 / 8  &235 & 18  / 11 &83   & 23 / 11 &--58  & 18 /10 \\
\hline
6280/4.00/--2.00   & \ha  & 74   & 10 / 13 &36    & 41 / 12 &--87   & 53 / 11 &290 & 27  / 24 &216  & 40 / 13 &69    & 42 /12 \\
                   & \hb  & 38   & 10 / 8  &--65  & 34 / 7  &--226  & 46 / 7  &280 & 27  / 10 &106  & 19 / 8  &--55  & 29 /9 \\
                   & \hg  & 34   & 15 / 7  &--103 & 34 / 6  &--288  & 47 / 7  &210 & 28  / 8  &73   & 20 / 7  &--106 & 29 /8 \\
\hline
6320/4.50/--2.00   & \ha  & 118  & 31 / 14 &44    & 60 / 13 &--110  & 60 / 11 &305 & 14  / 15 &183  & 40 / 13 &23    & 38 /12 \\
                   & \hb  & 57   & 22 / 8  &--66  & 42 / 7  &--230  & 49 / 7  &285 & 20  / 10 &91   & 16 / 8  &--68  & 26 /9 \\
                   & \hg  & 43   & 23 / 7  &--109 & 38 / 6  &--285  & 49 / 7  &285 & 31  / 9  &66   & 16 / 8  &--109 & 26 /8 \\
\hline
6550/4.50/--3.00   & \ha  & 109  & 25 / 13 &49    & 53 / 11 &--90   & 60 / 10 &233 & 31  / 14 &187  & 19 / 12 &46    & 38 /11 \\
                   & \hb  & 36   & 17 / 8  &--70  & 48 / 7  &--230  & 57 / 7  &232 & 25  / 9  &94   & 16 / 8  &--57  & 27 /8 \\
                   & \hg  & 26   & 20 / 7  &--119 & 36 / 6  &--297  & 54 / 6  &232 & 29  / 8  &72   & 17 / 7  &--103 & 28 /7\\
\hline
\end{tabular}%
\label{tab2}
\end{table*}


As a measure of the similarity, we used the root-mean-square deviation $\DIrms^2$ between the
normalized line profiles above a prescribed residual flux level, 

\beq
\DIrms^2 \equiv \frac{1}{N} \sum_{i=1}^N \left[f^\mathrm{3D}_i -
  f^\mathrm{1D}_i(\teff)\right]^2
\eeq

\noindent where $N$ is the number data points making up the line profile,
$f^\mathrm{3D}$ the flux based on the 3D, and $f^\mathrm{1D}$ the flux based
on the 1D model. We added explicitly the \teff-dependence in the 1D case as a
reminder that we varied the \teff\ of the 1D models in order to match the 3D
result. The 3D correction is obtained as the difference (in
the sense 3D-1D) between the effective temperature of the 3D model to the best
matching 1D one.

To calculate $\DIrms^2$ one has to define the portion 
of the profile over which this is computed. 
The line core must always be excluded in these
comparisons, since in real stars it is affected
by the presence of a chromosphere.
When fitting theoretical to observed spectra, the usual 
choice is to define a wavelength interval
which defines the wings of the line. Typical choices are 0.3\,nm and 0.5\,nm from
line centre \citep{cayrel1985}. In the comparison of theoretical spectra,
however, we may take advantage of the fact
that the continuum is perfectly 
defined and consider the wings as the portion 
of the profile that is above a given threshold.
In this way a similar fraction of the line wing is
considered, irrespective of the temperature of the
star. 
We believe this choice is the most appropriate
for the comparison of theoretical spectra, and use a threshold of 0.8 in
residual flux.  

In addition to the best matching \teff\ of a 1D model, the fitting provided further
information: i) the overall quality of the fit is related to the
residual $\DIrms^2$ at the best matching \teff; ii) the sensitivity to which we
can determine \teff\ is related to the rate of change of $\DIrms^2$ (more
precisely its curvature) with respect to \teff\ at the matching
point. Since we are dealing here with theoretical, essentially noise-less data,
the sensitivity is not a real issue, and we can fix the best matching \teff\
to arbitrary precision. However, in practice one is usually working with
spectra of only finite signal-to-noise ratio (S/N) so that the
sensitivity is important for the precision to which \teff\ can be obtained. To
make a connection to the practical limitations, we present
sensitivities and qualities of the fits in a form which makes it easy to relate
them to the situation one encounters in finite S/N spectra.

We assume a simple noise model where all pixels have the same S/N
distributed according to Gaussian statistics. This is well justified since
all pixels have a rather similar flux level due to our chosen high flux threshold
in the fitting. The standard deviation of the Gaussian distribution of the
flux $\sigF$ is given by $\sigF = (S/N)^{-1}$. Our minimization of $\DIrms^2$
would then transform to a \chisq-minimization. \chisq\ is related to $\DIrms^2$
as $\chisq = \frac{N}{\sigF^2} \DIrms^2$ 
where $N$ is the number of points which sample
the line profile.
The uncertainty of the fitted effective temperature~\sigT\ is given 
by (e.g., \citealt{Press})
\beq
\sigT=\left(\frac{1}{2}\ppdf{\chisq}{\teff}\right)^{-\frac{1}{2}} 
     = \sigF \sqrt{\frac{2}{N}} \left(\ppdf{\DIrms^2}{\teff}\right)^{-\frac{1}{2}}.
\label{e:sigT}
\eeq
\noindent Not surprisingly, the attainable precision scales inversely proportionally to
the S/N, and to $\sqrt{N}$. At this point, we perform
a more subtle re-interpretation of $N$: $\DIrms^2$ -- as an average -- does not
sensitively depend on $N$ as long as the line profile is sufficiently 
densely sampled. Hence, Eq.~\eref{e:sigT} does not only provide the scaling of
\sigT\ for the particular fit in question but for any fit based on a number of
$N$ points. We use this property to give \sigT\ always for the same nominal
S/N of 100 and number of points $N=700$ making all presented fits
inter-comparable. In the present context, \sigT\ is a measure of the
sensitivity -- or rather insensitivity -- of the Balmer line profiles to
changes in \teff. If one is dealing with spectra of finite S/N, \sigT\ provides
an estimate of the attainable precision, and Eq.~\eref{e:sigT} can be used to
translate the presented values to the actually present S/N and number $N$ of
statistically independent elements (pixels) in the spectrum under
consideration.

In a similar vein, we express the quality of the fit QF related to the
residual $ \DIrms^2$ at the fitting point in terms of a temperature
difference 
\beq
\mathrm{QF}= \sqrt{2\,\DIrms^2\mathrm{(residual)}}\,\left(\ppdf{\DIrms^2}{\teff}\right)^{-\frac{1}{2}}.
\label{e:QF}
\eeq
This provides a handy measure of whether the derived 3D \teff\ corrections make sense;
if the fit quality in terms of a temperature difference~QF is larger than the
derived correction~$\Delta$T one should take this as a warning that the 3D versus 1D
differences of the line profiles are so large that trying to match them
is not appropriate.

\section{Results}


For each 3D model, the 3D temperature correction ($\Delta T$), the quality of fit (QF)
and the sensitivity ($\sigma_{\rm Teff}$) for three different grids of LHD models
corresponding to three choices of \mlp\ are listed in Table \ref{tab2}.
The last nine columns of the table
provide the same quantities, but for the \mD\ model.  These allow us to
understand what part of the 3D temperature correction can be
ascribed to differences in the mean temperature profiles of a 3D model
and \xx\ model. Graphical representations of
the results are provided in the online appendix. 

The sample of investigated models is not large, however several patterns in
the temperature corrections clearly emerge.  At solar metallicity, \mlp =
0.5 provides nearly the same temperature correction for the first three
Balmer lines. This result matches what was derived from the comparison of
observed spectra to theoretical spectra based on 1D models by \citet{F93} and
\citet{1996A&A...309..879V}. We also find that the wings of H$\alpha$ are
relatively insensitive to \mlp. In the past this has motivated the use of only
H$\alpha$, but not of higher members of the series, for the determination of
\teff\ \citep{asplund,B07}. At low metallicity, however, the insensitivity is lost, and 
the temperature correction of H$\alpha$ is not independent of \mlp.  
There are also sizeable differences in the temperature corrections of the
various Balmer lines. 

In Table~\ref{tab3} we provide the temperature correction spread
$\delta$T, defined as $\Delta$T$_\mathrm{max} -\Delta$T$_\mathrm{min}$ among
the first three lines of the Balmer series.  For the models at \teff =5500\,K
and \teff = 5920\,K the smallest spread is achieved for \mlp=2.0. 
For the \teff = 5500\,K model the spread in the
temperature correction is not even a monotonic function of \mlp.  However, the spread
for \mlp=2.0 is not considerably smaller than that for \mlp = 0.5, especially
when compared to the associated errors.  This is not surprising; the
mixing-length theory is a parametric phenomenological description of
convection and it is obvious that it is not possible for a single value of the
free parameter to capture all the complexity of this physical phenomenon.  It
can be seen that the situation is more complicated for the cooler models,
while for the three models hotter than 6000\, K, $\delta$T is a monotonic
value of \mlp\ and achieves the smallest value for \mlp = 0.5.

The $\mD - \xx$ differences appear to be quite regular: the lower the
temperature, the higher the temperature correction, reaching the value of 500\,K for
the model at \teff=5500\,K.  This regularity is not immediately obvious when
looking at the $3\mathrm{D} - \xx$ differences, since these depend also on the
temperature fluctuation.  The importance of temperature fluctuations is
different for the different \teff, but also for different metallicities and
gravities.  In all cases we see that the role of the temperature fluctuations
is to {\em reduce} the temperature correction, with respect to what is derived with
respect to the \mD\ model.  But this reduction can be as large as 50\% (for
the \teff=5500\,K model) or as small as 7\% (for the \teff=5920\,K model).

\begin{table}
\centering
\caption{Spread of $\Delta$T among Balmer series members.}
\begin{tabular}{lllrrrrrr}
\hline
\teff & \logg & [M/H]  &  \multicolumn{3}{c}{$\delta$T(3D -- 1D$_{\rm LHD}$)}  & \multicolumn{3}{c} { $\delta$T($\left\langle\mathrm{3D}\right\rangle$ -- 1D$_{\rm LHD}$)} \\
  K   &  c.g.s&  dex   &  \multicolumn{3}{c}{K} &\multicolumn{3}{c}{K}\\
 \hline
      \multicolumn{3}{c}{\mlp}      &    {0.5} &  {1.0} &  {2.0} &  {0.5} &  {1.0} & {2.0} \\
 \hline \hline
5500  & 3.50  & --2.00   & 245   & 343& 233  & 280 & 251 & 202 \\ 
5780  & 4.44  & \, 0.00  &  5    & 88 & 242  & 129 & 32  & 104  \\
5920  & 4.50  & --3.00   &  180  & 151 & 90  & 70  & 135 & 79   \\
6280  & 4.00  & --2.00   &   40  & 139 & 201 & 80  & 143 & 175  \\
6320  & 4.50  & --2.00   &  75   & 153 & 175 & 20  & 117 & 132  \\
6550  & 4.50  & --3.00   &  86   & 168 & 207 & 1   & 115 & 149 \\
\hline
\end{tabular}%
\label{tab3}
\end{table}

\section{Conclusions}

The main conclusion that can be drawn from this investigation is that stellar
granulation has sizeable effects on the wings of the Balmer lines.  When
quantified in terms of the temperature correction, such differences span
the range of $\pm 300$\,K for the investigated models.  This implies
that if high accuracy effective temperatures are to be derived from the wings
of Balmer lines, the effects of granulation must be taken into account.  A
temperature scale based on fitting the wings of Balmer lines with 1D models
will be different from that derived by using 3D models.  The difference
between the two scales is not a simple offset, but rather has a temperature
dependent slope.


The smallest spread in temperature correction often occurs when $\mlp=0.5$, but not
always. The temperature correction for \ha\ is generally larger than that of the
other members of the series.
At lower temperatures the sensitivity of \ha\ to effective temperature drops.
This is illustrated by our cooler model (\teff = 5500\,K); remarkably,
the higher members of the series maintain a rather good sensitivity to \teff.
The use of 3D models, removing the uncertainty on the choice of \mlp,
suggests that the use of several Balmer lines should greatly increase the
accuracy and robustness of the \teff\ determination -- provided one can handle
the increasing line blending for the higher series members.

While a bias of the true \teff\ of a late-type star determined from 1D
  models seems unavoidable, the question occurs as to whether a \teff\ corrected
  for the 3D-1D difference is really superior if one wants to use the underlying 1D
  model for the interpretation of other features in the stellar spectrum --
  typically spectral lines for abundance determinations.  Arguably, the
  calibration of the 1D model inherent to the temperature fitting can make it
  advantageous to rather use the uncorrected 1D temperature.  However, this
  hinges on the relation between the spectral feature of interest and the
  one fitted for the \teff\ determination, and has to be decided upon
  on a case by case basis. In the online appendix we present a case where a
  \teff-corrected 1D model performs better.
 
The general trends highlighted in our investigation need to be confirmed by
the use of a larger grid of 3D models.  In addition the implementation of a
more up-to-date line broadening theory will allow a direct comparison between 3D
synthesis and observed spectra.  In the near future we plan to extend our
investigation in these two directions.

\begin{acknowledgements}
  HGL, NTB, and PB acknowledge support from EU contract MEXT-CT-2004-014265
(CIFIST).
We thank the supercomputing center CINECA which has granted us time to compute
part of the hydrodynamical models used in this investigation, through
the INAF-CINECA agreement 2006, 2007.
\end{acknowledgements}

\begin{appendix}

\onecolumn

\section{Graphical representation of the 3D temperature correction, $\Delta$T,
  the quality of the fit, QF, and the sensitivity of the fit, $\sigma_{\rm Teff}$}

\begin{figure*}
\begin{center}
\includegraphics[width=1.00\textwidth]{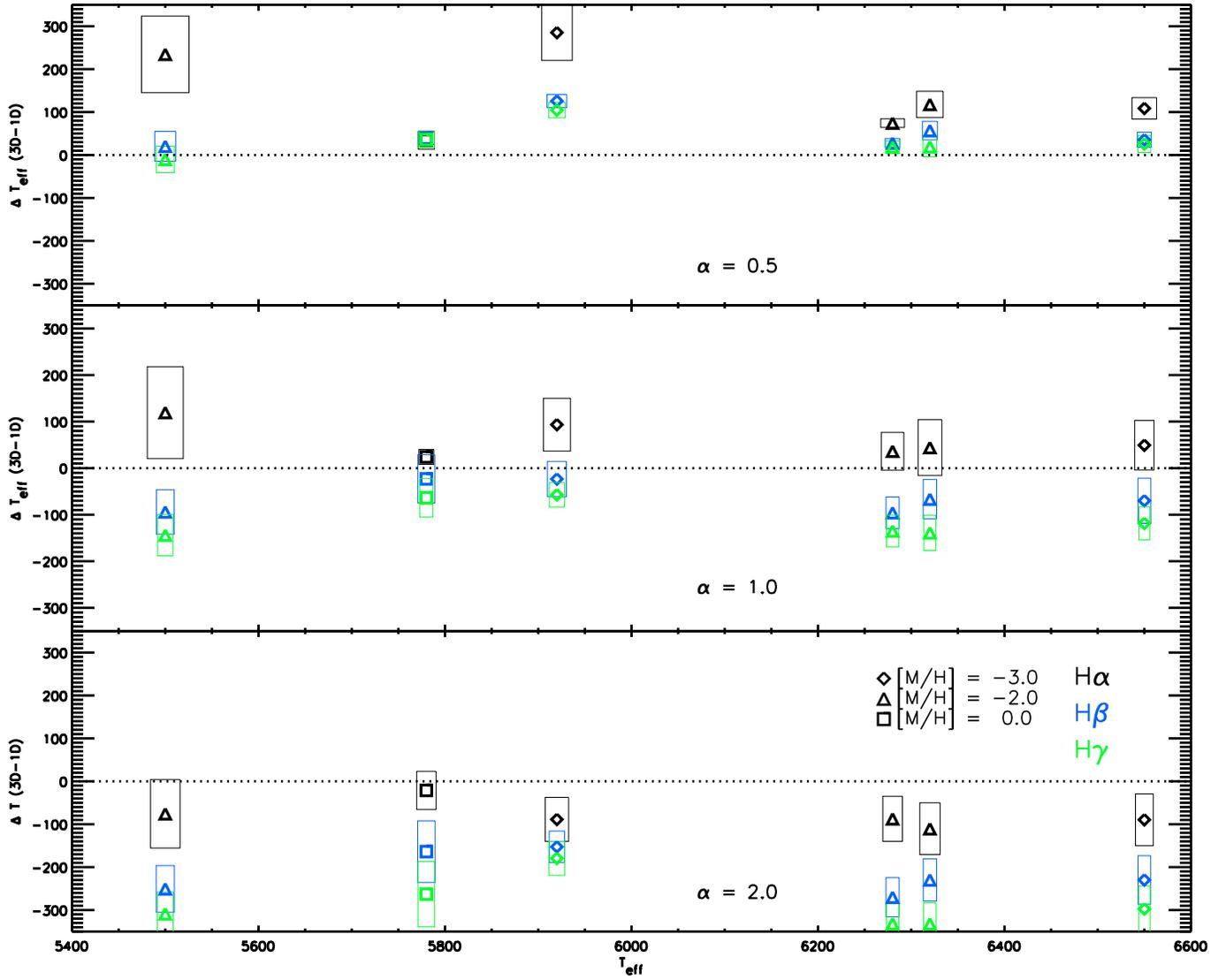}
\end{center}
\caption[]{
The temperature differences (3D -- 1D) using 1D models with
$\alpha$ = 0.5, $\alpha$ = 1.0 and $\alpha$ = 2.0, for H$\alpha$, H$\beta$,
and H$\gamma$ for all of the six models considered. The uncertainties related
to each temperature difference measurement are displayed as boxes. The value of the
uncertainty on the $x$-axis corresponds to the associated $\sigma_{\rm Teff}$ (sensitivity) values, while
the uncertainty on the $y$-axis corresponds to the associated QF (quality of fit)
values. See Table~\ref{tab2} and the text for details.
}
\label{fig2}
\end{figure*}

\twocolumn

\section{3D-1D temperature correction and the lithium abundance in metal-poor F-type dwarfs}

\begin{figure}
\begin{center}
\includegraphics[width=\columnwidth]{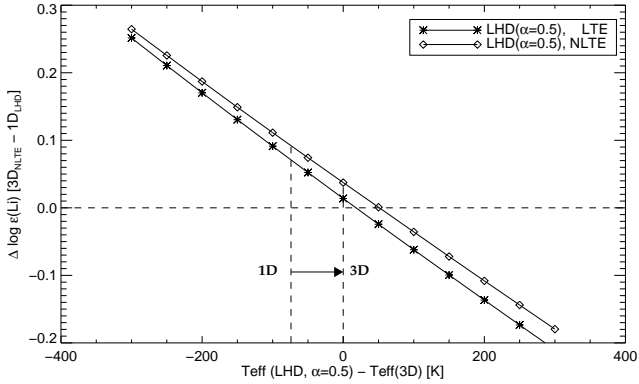}
\end{center}
\caption[]{
3D-1D abundance correction versus effective temperature difference for the
test case of a metal-poor F-type dwarf. For details see text.}
\label{fig3}
\end{figure}

Here we present an example where it is advantageous to use a 1D model
corrected for the 3D-1D temperature difference when deriving the chemical
abundance from spectral line analysis: the abundance of lithium in a
metal-poor F-type dwarf obtained from the 670.7\,nm resonance line. We
performed a 3D-NLTE spectrum synthesis calculation for the line on the 3D
model with \teff=6280\,K, \logg=4.0, [M/H]=-2.0 (cf. Table~\ref{tabmod}).  We
considered the resulting spectrum as representing an observation. Unlike a
real observation, however, the underlying lithium abundance and stellar parameters
are exactly known.  According to Table~\ref{tab3} the temperature correction
from H$\alpha$ fitting amounts to 74\,K in this case, i.e. the 1D model that
fits the 3D H$\alpha$ profile best is 74\,K cooler than the 3D model. We then
calculated for a series of thirteen 1D (LHD) models of different effective
temperatures 1D lithium line profiles in LTE and NLTE. In the 1D spectrum synthesis,
we assumed a microturbulence velocity of 1\,km/s; however, the actual value is
not important since the line was chosen to be very weak. We derived for each
model the lithium abundance matching the line strength obtained in 3D.
Figure~\ref{fig3} depicts the resulting abundance differences between 
the underlying lithium abundance assumed in the 3D model and the derived
1D abundance, versus the effective
temperature differences between 1D models and the 3D model. As evident from the
plot, one reduces the abundance error resulting from the erroneous effective
temperature of the 1D model when applying the 3D-1D temperature correction.
This holds irrespective of whether the 1D abundance analysis is performed in LTE
or NLTE. Not surprisingly, the figure also shows that the correction of the
effective temperature does not result in a perfect match of the lithium
abundances in a 1D analysis.

\end{appendix}

\end{document}